\documentclass[twocolumn, superscriptaddress, nofootinbib]{revtex4-2}

\usepackage{graphicx}% Include figure files
\usepackage{dcolumn}% Align table columns on decimal point
\usepackage{bm}% bold math
\usepackage{lipsum}
\usepackage{amsmath}
\usepackage{gensymb}
\usepackage{float}
\usepackage{amsmath}
\usepackage{amssymb}
\usepackage{physics}

\begin{document}

\preprint{APS/123-QED}

\title{Polarization-Sensitive Diffractive Optics and Metasurfaces: ``Past is Prologue''}

\author{Noah A. Rubin}\email{noahrubin@ucsd.edu}
\affiliation{Department of Electrical and Computer Engineering, University of California San Diego, La Jolla, CA 92093 USA}
\author{Yeshaiahu Fainman}
\affiliation{Department of Electrical and Computer Engineering, University of California San Diego, La Jolla, CA 92093 USA}

\date{\today}% It is always \today, today,

\begin{abstract}
Polarization control and switchability are among the most unique features of ``metasurfaces'' as compared with diffractive optics technologies of the past. Here, we review how the polarization control afforded by the advent of present-day metasurfaces compares to diffractive elements of previous decades, clarifying from a functional perspective what is new, and what is not.
\end{abstract}

\maketitle

\section{Introduction} 
\vspace*{-3pt}

Metasurfaces -- subwavelength-structured diffractive optical elements -- have attracted significant attention in the optics and photonics research communities over the last decade. This field is often cited for its promise to miniaturize optical components such as lenses, filters, and other free-space optical components with ``flat optics'' that can be manufactured using broadly scalable semiconductor fabrication techniques. However, under the guise of different labels (in particular, ``holography'' and ``diffractive optics'') many of these concepts have a long past history of development.

Nonetheless, we contend that the space-variant polarization control offered by metasurfaces distinguishes them from past work. We note that this perspective is shared even by some of the greatest skeptics of metasurface research generally \cite{Lalanne2017, Banerji2019} (a point of commonality even if the reasons for this skepticism differ greatly). Today, functionally new polarization optical components are being realized with metasurface-like components -- contributing to both optics research and generating new technologies. However, as with most developments in nanophotonics research \cite{Lalanne2023}, polarization-sensitive diffractive optical elements are not themselves wholly new. Thus, it is important to approach this with nuance, simultaneously acknowledging the novelty of metasurface polarization optics with a keen awareness of where these do (and \emph{do not}) represent an advance over the ideas and devices of the past.

As it so happens, the development of designer optical elements that permit space-variant control of polarized light has an especially long history (going back to at least the 1960s). This is a story that spans decades, countries, and technological platforms --- from holographic materials in film, to liquid crystals, as well as solid state components prepared using standard semiconductor manufacturing techniques. This latter category (of which the ``metasurfaces'' of today are a part) will be our focus here. However, one of us (NAR) recently authored a comprehensive review article \cite{Rubin2021a} discussing all of this at length. Our discussion that follows is a briefer account of content contained therein.

In what follows, we compare contemporary metasurface polarization optical components to those of decades past. Our perspective as authors is shaped by one of us having been involved in this field in the last several years (NAR), with the other having been active in the early 1990s (YF). This 30 year gap permits clarity on what has changed, and what has not.

In Sec. \ref{sec:math}, we introduce a mathematically rigorous means of comparing the different polarization-sensitive diffractive optics technologies. This culminates in a simple mathematical visualization we term \emph{retarder space} -- past technologies differ in the extent of this retarder space they can access. In Sec. \ref{sec:comparison}, we briefly introduce three polarization-sensitive diffractive optics technologies of decades past, classify them within this retarder space, and show how present day metasurface-like elements occupy a larger subspace within it. In Sec. \ref{sec:implications_and_conclusion}, we discuss the implications of this higher degree of control (both in terms of specific devices which have been realized and research directions created) and conclude with some possible future directions.

\section{Understanding Space-Variant Diffractive Optics}
\label{sec:math}

In this perspective, we will compare metasurface polarization optics to several past demonstrations of polarization sensitive diffractive optical elements which, like metasurfaces, were also manufactured using lithographic, deposition, and etching techniques. Our comparison here will be on the basis of \emph{optical function} (that is, the extent of space-variant polarization control enabled -- as defined mathematically below). We note that purely \emph{technological} differences (e.g., different materials, wavelengths, or -- most significantly -- the availability of improved fabrication techniques/equipment) shall not interest us here even if, perhaps, these are the true enablers of present-day metasurfaces.

However, we first develop a rigorous, mathematical basis on which to compare different polarization-sensitive diffractive optical elements.

\subsection{Classifying space-variant polarization control}

We consider here optical elements which implement a local polarization transformation on an incident wavefront, i.e.,

\begin{equation}
\label{eq:local_pol_trans}
    \ket{E^{\prime}(x^{\prime},y^{\prime})} = \boldsymbol{J}(x^{\prime},y^{\prime}) \ket{E(x^{\prime},y^{\prime})}.
\end{equation}

In Eq. \ref{eq:local_pol_trans}, bra/ket notation is used to denote two-element Jones vectors (such as $\ket{E(x^{\prime},y^{\prime})}$); the (coherent) interaction of these with an optical element which can modify polarization is given by a $2\times2$ Jones matrix $\boldsymbol{J}$. Each element of Eq. \ref{eq:local_pol_trans} is allowed to be space-variant with transverse coordinates $(x^{\prime},y^{\prime})$\footnote{Throughout this manuscript, coordinates with $^\prime$ attached are intended to denote Cartesian spatial coordinates -- this is to properly distinguish the $x^{\prime}$ and $y^{\prime}$ spatial coordinates from $x$ and $y$ linearly polarized light along these same coordinate axes. This distinction is important in equations that follow, such as, e.g., in Eq. \ref{eq:pol_control_fbh} where labels for $x$ polarized light and the $x^{\prime}$ spatial coordinate appear together and must be distinguished.}. This includes $\boldsymbol{J}(x^{\prime},y^{\prime})$ -- which is then a mathematical description of the transfer function enabled by a polarization-sensitive diffractive optical component. Inherent in the description implied by Eq. \ref{eq:local_pol_trans} is the locality of the polarization transfer function. Or, in other words, Eq. \ref{eq:local_pol_trans} is a statement that the element implementing $\boldsymbol{J}$ is \emph{thin} (relative to the wavelength)\footnote{This condition is important so that this most simple amplitude transmittance model is valid; otherwise, were the element being described ``thick'' enough, diffraction could occur inside the optical element itself such that the locality of Eq. \ref{eq:local_pol_trans} would be an inappropriate description. The rigorous distinction between ``thin'' and ``thick'' diffractive optical elements is a subtle one \cite{Gaylord1981, Gaylord1985} which is outside of our immediate scope here.} and moreover that all spatial frequencies involved (either due to steeply off-normal incidence or rapid variation in $\boldsymbol{J}(x^{\prime},y^{\prime})$ near the wavelength scale) are small such that the longitudinal component of the electric field along $z$ can be neglected.

Any set of four complex numbers can form a valid Jones matrix \cite{Chipman2019} -- thus, a Jones matrix contains eight free parameters (corresponding to the real and imaginary parts of each matrix element). However, any given technology realizing $\boldsymbol{J}(x^{\prime},y^{\prime})$ may only be able to implement some subset of these freely from point-to-point across the $(x^{\prime}, y^{\prime})$ plane of an optical element. The extent of control over the Jones matrix transformation is thus a logical classification scheme for polarization-sensitive diffractive optics \cite{Rubin2021a}.

This has a direct parallel in \emph{scalar} diffractive optics, in which the transfer function is a single space-variant complex number $t(x^{\prime},y^{\prime})$. Certain diffractive components (gratings, SLMs, etc.) are often referred to as phase- or amplitude-only, and within these, different sorts of phase or amplitude control can be realized by different approaches (e.g., binary phase/amplitude, or perhaps limited continuous phase control over a $\pi$ range). This enables simple comparisons between technologies.

The simple division of a complex number into amplitude and phase -- known as the scalar polar decomposition, widely used to classify diffractive elements -- can be extended in the polarization-dependent case using the \emph{matrix} polar decomposition, whereby any Jones matrix $\boldsymbol{J}$ can be written as

\begin{equation}
\label{eq:polar_decomposition}
    \boldsymbol{J} = \boldsymbol{H}\boldsymbol{U}.
\end{equation}

That is, any Jones matrix $\boldsymbol{J}$ can be written as the product of a Hermitian Jones matrix $\boldsymbol{H}$ (with $\boldsymbol{H}^{\dagger} = \boldsymbol{H}$; $\dagger$ is the Hermitian conjugate) and a unitary Jones matrix $\boldsymbol{U}$ (with $\boldsymbol{U}^{-1} = \boldsymbol{U}^{\dagger}$) \cite{Chipman2019, Rubin2021a}. Each of $\boldsymbol{H}$ and $\boldsymbol{U}$ has four degrees of freedom defined by angular quantities. The polar decomposition has a simple physical interpretation, in terms of polarization elements -- unitary Jones matrices encode lossless waveplate-like transformations, and Hermitian Jones matrices encode amplitude-modulating polarizer-like behavior. The former is a generalization of ``phase'' to the higher-dimensional space of polarization optics, while the latter is a vectorial generalization of amplitude. With the aide of Eq. \ref{eq:polar_decomposition}, any polarization transformation can be understood as a cascade of a waveplate retarder followed by a (partial) polarizer.

In diffractive optics, phase-only elements are usually desired since phase shifts are energy-preserving. After diffraction to the far-field, phase-only diffractive optics can implement amplitude modulated patterns by spatially redistributing energy rather than selectively absorbing it. For similar reasons, almost all work in polarization-sensitive diffractive optics has focused on engineering unitary transformations. We thus restrict the discussion that follows to cases in which $\boldsymbol{J}(x^{\prime},y^{\prime})$ is everywhere unitary, i.e., cases in which $\boldsymbol{H} = \mathbb{I}$ where $\mathbb{I}$ is the $2\times2$ identity matrix.

An optical element implementing a strictly unitary $\boldsymbol{J}(x^{\prime},y^{\prime})$ has four degrees-of-freedom at each point $(x^{\prime},y^{\prime})$. In particular, any unitary Jones matrix can be written as

\begin{equation}
\label{eq:basis_vectors}
    \boldsymbol{U} = e^{i\phi}\big(e^{i\frac{\Delta}{2}}\ketbra{u}{u} + e^{-i\frac{\Delta}{2}}\ketbra{u^{\perp}}{u^{\perp}} \big).
\end{equation}

Here, $\ket{u}$ and $\ket{u^{\perp}}$ are an orthonormal basis of polarization states (Jones vectors), such that $\braket{u} = 1$, $\braket{u^{\perp}} = 1$, and $\braket{u}{u^{\perp}} = 0$. The outer products $\ketbra{u}{u}$ and $\ketbra{u^{\perp}}{u^{\perp}}$ in Eq. \ref{eq:basis_vectors} are $2\times2$ matrices\footnote{To be clear about the bra-ket notation used here, if $\ket{u}=\begin{pmatrix} \cos\chi \\ \sin\chi e^{i\delta} \end{pmatrix}$, then $\ketbra{u}{u} = \begin{pmatrix}
\cos^2\chi & e^{-i\delta}\cos\chi\sin\chi  \\ e^{i\delta}\cos\chi\sin\chi & \sin^2\chi
\end{pmatrix}$.}. This orthonormal basis can be represented by two angles $\chi$ and $\delta$ such that

\begin{equation}
\label{eq:normalized_jones}
    \ket{u} = \begin{pmatrix} \cos\chi \\ \sin\chi e^{i\delta} \end{pmatrix}, \ket{u^{\perp}} = \begin{pmatrix} -\sin\chi \\ \cos\chi e^{i\delta} \end{pmatrix}
\end{equation}

and connotes the eigenbasis of the unitary matrix, that is, the basis of polarization states delayed by $\emph{U}$. In the case where the eigenbasis consists of linear polarization states (i.e., if $\delta=0$), Eq. \ref{eq:basis_vectors} describes the Jones matrix of a waveplate, and the azimuthal angle of these basis vectors gives the orientation of the waveplate's fast axis.

Finally, in Eq. \ref{eq:basis_vectors} $\phi$ represents a polarization-independent, overall phase shift, while $\Delta$ represents the element's \emph{retardance}, the differential phase accrued by the eigenbasis $\ket{u}/\ket{u^{\perp}}$ which is bounded in the range $\Delta \in (-\pi, \pi]$.

Any unitary polarization transforming element is uniquely specified by the four angles $\phi$, $\Delta$, $\chi$, and $\delta$. In a polarization-sensitive diffractive optical element (such as a metasurface), these angles can vary with space. However, a given technology for implementing these transformations \emph{cannot} in general access the entire four-dimensional subspace spanned by these angles, or, that is, in general only a \emph{subset} of the broadest set of possible unitary matrix transformations can be synthesized by a given diffractive optical element from point-to-point across its aperture.

This subset of polarization transformations attainable by a component provides a rigorous way to compare the function of different polarization-sensitive, diffractive optics technologies.

\subsection{Visualizing Retarder Space: Pictorial Comparisons}
\label{subsec:retarder_space}

\begin{figure}
    \centering
    \includegraphics[width=\linewidth]{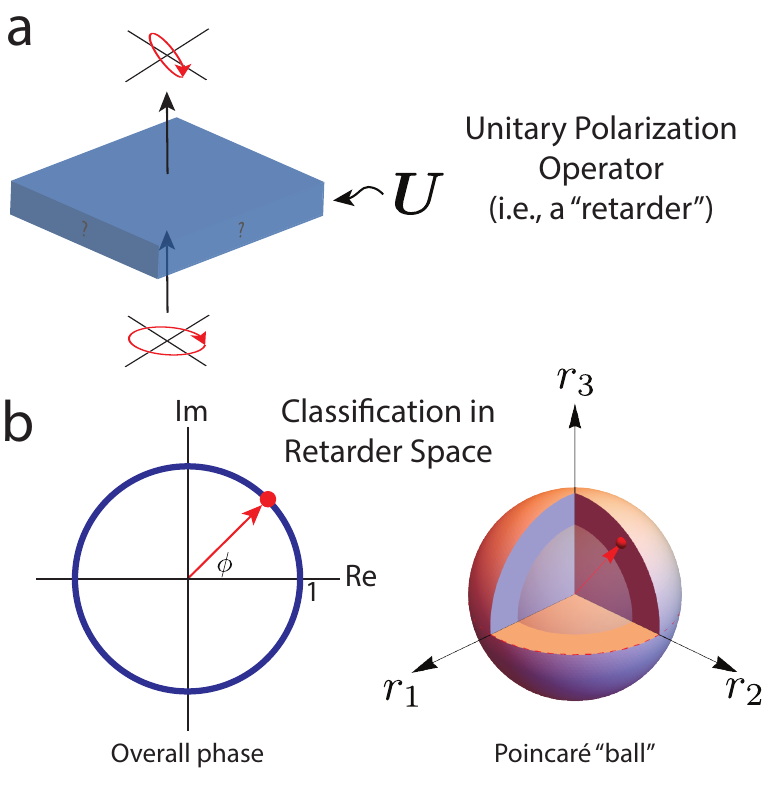}
    \caption{A unitary polarization transformation (that is, a phase retarder) given by Jones matrix $\boldsymbol{U}$ can be classified and visualized in ``retarder space''. The four degrees of freedom of matrix $\boldsymbol{U}$ can be visualized as, first, a point on the unit circle (connoting the matrix's overall phase $\phi$ and as a point within the 3D volume of the Poincar\'e ``ball''. The radius within the ball is governed by $\boldsymbol{U}$'s retardance, and the direction of the point from the origin corresponds to the Stokes vectors of $\boldsymbol{U}$'s eigen-polarization states on the ordinary Poincar\'e sphere.}
    \vspace{-0.5cm}
    \label{fig:fig1}
\end{figure}

In scalar diffractive optics, where the transfer function is given by a single space-variant complex number $t$, the two-dimensional complex plane is a useful and complete visualization tool. A similar (albeit higher-dimensional) visualization can be developed here to understand the space occupied by unitary Jones matrix transfer functions (as defined in Eq. \ref{eq:basis_vectors}).

This visualization is known as \emph{retarder space} \cite{Rubin2021a, Chipman2019}. First, recall that Jones vectors (such as the eigenbasis defined in Eq. \ref{eq:normalized_jones}) can be visualized on the Poincar\'e sphere, wherein the poles represent circular polarization states of opposite handedness, the equator represents linear polarization states, and the two hemispheres are elliptical polarization states in general. The first component of retarder space is a solid Poincar\'e sphere (a ball, so to speak) with unit radius (Fig. \ref{fig:fig1}(b), right). The axes of this Poincar\'e ``ball'' (a nonstandard term used for clarity here) are denoted as $(r_1, r_2, r_3)$ and are analogous to the usual Stokes vector components, but differently labeled to stress that this sphere is a space of polarization operators (Jones matrices) rather than polarization states themselves (as would be given by Jones or Stokes vectors). The eigen-polarization states of the unitary matrix (as given in Eq. \ref{eq:basis_vectors}) define a direction within this sphere. In particular, the Stokes vector corresponding to the Jones vectors Eq. \ref{eq:basis_vectors} (found with aide of the Pauli matrices, see \cite{Rubin2021a}, Sec. 2.5) defines a direction from the origin of the sphere. The radius from the origin is given by $\sin\frac{\Delta}{2}$ (defined in this way to bound the radius to unity for $\Delta=\pi/2$, with $\Delta$ again being the retardance). Together, then, the parameters $\chi$, $\delta$, and $\Delta$ define a unique point within the unit-radius spherical volume of the Poincar\'e ``ball''. Entire classes of polarization optical components can be visualized as sections of this spherical space -- for instance, the origin represents fully polarization-independent transformations (i.e., $\boldsymbol{U} = \mathbb{I}$), while a shell centered at the origin located at a radius of $\sqrt{2}/2$ (i.e., $\Delta = \pi/2$) represents the set of all possible quarter-wave retarders.

This visualization is missing only the overall phase $\phi$. The second component of retarder space, then, is a unit circle on which this overall phase shift can be visualized as an angle (Fig. \ref{fig:fig1}(b), left).

Any unitary Jones matrix $\boldsymbol{U}$ can then be uniquely specified by a point in the volume of the Poincar\'e ball (determined by $\chi$, $\delta$, and $\Delta$) and a point on the unit circle (which determines $\phi$) as in Fig. \ref{fig:fig1}(b). A \emph{complete} polarization-sensitive diffractive optical element could implement any point in the entire volume of this unit sphere and any point on the perimeter of the unit circle independently and arbitrarily across its spatial aperture. The subspace of this which any given technology can access can be used as a simple means of functional comparison.

\section{Past vs. Present}
\label{sec:comparison}

In this section, we briefly introduce three approaches to polarization-sensitive diffractive optical elements which formed the basis of distinct research fields in the 1990s and early 2000s. The function of these is detailed using the retarder space previously introduced. Current generation metasurface polarization optics are shown to enable a higher level of functionality, i.e., a greater subspace of the most general retarder space defined above, than these past approaches.

Each technology discussed in this section forms a column of Fig. \ref{fig:fig2}, in which the technology approach is illustrated alongside the coverage of the overall retarder space it provides.

\subsection{``Multi-level polarization-selective computer generated holograms''}
\label{subsec:fbh_cgh}

In the early 1990s, Fainman et al. developed polarization-selective computer generated holograms \cite{Ford1993, Xu1995, Xu1996a, Krishnamoorthy1997}. Here, polarization-selectivity is provided by the use of birefringent substrate material (LiNbO$_3$ or calcite, in this case), with multi-level lithography and etching used to define propagation depth in a spatially-variant manner. (This is reviewed in more detail in \cite{Rubin2021a}, Sec. 4.3).

In essence, these elements permit the realization of a switchable phase hologram, implementing independent phase profiles for incident $x$ or $y$ polarized light (where $x$/$y$ is defined relative to the in-plane fast and slow axes of the substrate used, distinguished from the spatial coordinates $x^\prime/y^\prime$). In other words, these optics implement a space-variant, unitary Jones matrix of the form

\begin{equation}
\label{eq:pol_control_fbh}
    \boldsymbol{J}(x^{\prime},y^{\prime}) = \begin{pmatrix} e^{i\phi_x (x^{\prime},y^{\prime})} & 0 \\ 0 & e^{i\phi_y(x^{\prime},y^{\prime})}\end{pmatrix},
\end{equation}

wherein $\phi_x(x^{\prime}, y^{\prime})$ and $\phi_y(x^{\prime}, y^{\prime})$ are the independent, space-variant phase profiles imparted on $x$ and $y$ polarized light, respectively, as a function of transverse spatial coordinates $(x^{\prime}, y^{\prime})$. This capability was shown to be capable of realizing a variety of polarization switchable behaviors in \cite{Ford1993, Xu1995, Xu1996a, Krishnamoorthy1997}.

As shown in Fig. \ref{fig:fig2} (leftmost column), as visualized in retarder space, polarization control of the form of Eq. \ref{eq:pol_control_fbh} can implement independent and arbitrary control of retardance $\Delta$ and overall phase $\phi$. The axis of this retardation is, however, fixed across the device (along the local $x^{\prime}$/$y^{\prime}$) by the fast/slow axes of the birefringent substrate medium which are uniform across the device aperture. Moreover, the extent of phase and retardance control is discrete, being defined by the number of aligned lithography/etching steps undertaken during fabrication. In retarder space, this is visualized as discrete points along the unit circle (corresponding to discrete overall phase control $\phi = (\phi_x + \phi_y)/2$) and discrete points along the $r_1$ axis of the Poincar\'e ball (spaced at radii according to the retardance $\Delta = \phi_x - \phi_y$). At the bottom of the first column of Fig. \ref{fig:fig2}, the discrete phase control enabled by these multi-level phase elements is depicted by discrete points in the complex plane (shown here for the specific case of 8-level phase control).t

This work was one of the earliest solid-state implementations of polarization-sensitive diffractive optical elements and influenced a host of future works (including in metasurfaces). It presented several drawbacks, however. For one, the use of multi-level phase control is inherently challenging -- in this case, $N$ level phase control on $\phi_x$ and $\phi_y$ would require $4\log_2 N$ lithography/etch steps to achieve, as well as an onerous alignment procedure since two substrates were generally used \cite{Rubin2021a}).

\subsection{``Form-birefringent elements''}

\begin{figure*}
    \centering
    \includegraphics[width=\textwidth]{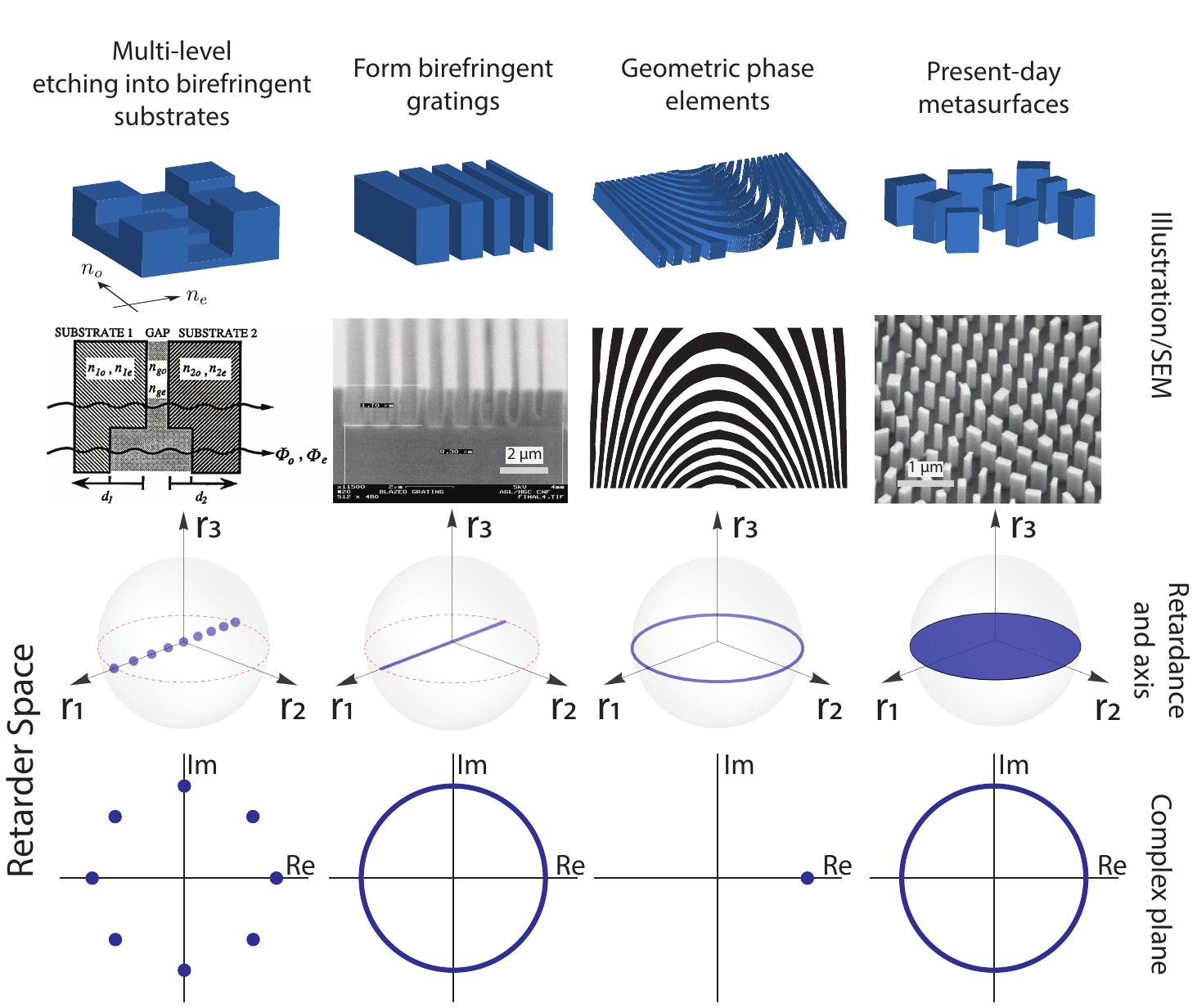}
    \caption{Classification of previous generations of polarization-sensitive diffractive optics and present-day metasurface-like elements. Each column represents a different technology. The top two rows depict each technology (using SEMs where possible), while the bottom two rows display the extent of coverage these offer in the retarder space defined in Sec. \ref{subsec:retarder_space} and Fig. \ref{fig:fig1} -- using both the Poincar\'e ball and the unit circle to depict overall phase. In the illustration row, the leftmost two images have been adapted from \cite{Ford1993, Lopez2001} © Optical Society of America.}
    \label{fig:fig2}
\end{figure*}

In part due to these consideration, several groups such Fainman et al. \cite{Richter1995, Xu1996, Tyan1997}, Craighead et al. \cite{Lopez1998, Lopez2001}, and Aoyama et al. \cite{Aoyama1991}, instead developed \emph{form-birefringent} polarization-sensitive diffractive optical elements in the mid-1990s (discussions under the name ``form birefringence'', however, go back to early editions of Born \& Wolf \cite{Born1999}). Much was written about the physics of this effect in these references (summarized in Sec. 4.7c of \cite{Rubin2021a}), but the essence of the effect is that sub-wavelength ridges in a dielectric medium are experienced differently by light polarized along the ridge versus perpendicular to the ridge. In transmission, light then experiences a polarization-dependent phase shift which can be well approximated by an ``effective medium'' picture \cite{Rytov1956} which predicts a form birefringence between orthogonal polarization (that is, birefringence due to the physical form of a structure, rather than inherent to its material). An example of these structures is given in Fig. \ref{fig:fig2} (second column from the left).

Mathematically, the polarization control afforded by these devices is similar to that of Eq. \ref{eq:pol_control_fbh} with several significant advantages. For one, these form-birefringent devices could be made with a single lithographic/etch step, moreover without the need to align holograms written on separate substrates. Second, the level of phase control is continuous rather than discrete -- changes in the width of the ridges adjust the birefringence experienced in a continuous manner. Thus, as viewed in retardance space in Fig. \ref{fig:fig2}, the control afforded by these form birefringent components is similar to that of Sec. \ref{subsec:fbh_cgh} (except continuous rather than discrete) --- the fast axis of the spatially-varying retarder is constrained to lie along a single direction ($x^{\prime}/y^{\prime}$), with control over the retardance and overall phase. An important limitation here, however, is that with these form birefringent elements there is only one degree of freedom - that being the width of the ridge (or equivalently, its duty cycle in a periodic arrangement of ridges). Changing the width of the ridge affects both $\phi_x$ and $\phi_y$ simultaneously in a way that is impossible to disambiguate -- that is, both birefringence $\Delta$ and $\phi$ are altered and cannot be independently specified.

We note, however, that much work in this area was very simple in function (at least viewed in hindsight). Often these form-birefringent devices implemented only spatially-uniform polarization behavior, usually with the goal of creating polarization beamsplitting behavior in reflection and transmission \cite{Tyan1997, Lopez1998}. A few of these works, however, instead acted as polarization beamsplitters by implementing separate phase masks for orthogonal linear polarizations, directing these light to different diffraction orders based on its polarization \cite{Aoyama1991, Xu1996}, closer to the spirit of the polarization-sensitive diffractive optics reviewed in this article. However, this phase control was binary in nature. So, while the diffractive optical elements of this era were in theory capable of the more general continuous phase/birefringence control depicted in the second column of Fig. \ref{fig:fig2}, we are not aware of any works that actually used this in its full generality\footnote{In fact, the SEM shown in this second column is drawn from a work in which polarization-selective behavior was not even explicitly intended \cite{Lopez2001} --- we find it, however, to be the highest-quality, most exemplary SEM from this period to demonstrate communicate the historical perspective here (most of the other works cited in this section lack clear electron micrographs).}.

Importantly, form birefringence is the mechanism of polarization-dependent phase control used by contemporary metasurface elements. This line of work was thus historically significant and foreshadowed later research.

\subsection{``Space-variant Pancharatnam-Berry Phase elements''}
\label{subsec:geometric_phase}

These early developments in form birefringence enabled the creation of a second class of polarization-selective diffractive optical elements in the early 2000s. These elements exploited the Pancharatnam-Berry phase, and formed the basis of a research effort from Hasman et al. (There are too many references to cite here, so we note their summary review article \cite{Hasman2005}; we also note \cite{Levy2004, Tsai2006} from some of the same researchers who investigated the form birefringent nanostructures of the previous section).

The Pancharatnam-Berry phase is a geometric phase effect that comes about from polarization change -- in its simplest manifestation, circular polarization passing through a half-waveplate changes from, e.g., right-handed to left-handed, but accrues a phase shift equal to twice the fast-axis angle of the waveplate. In this way, a form-birefringent grating element whose width is selected to impart a halfwave birefringence (that is, $\phi_x - \phi_y = \pi$) can be rotated to impart any desired phase profile on one circular polarization (say, right-handed) as $\phi(x^{\prime},y^{\prime}) = 2\theta(x^{\prime},y^{\prime})$ where $\theta$ is the space-variant orientation of the ridges' long axis.

In most of these works, the polarization-dependence of these devices is treated as an afterthought -- the geometric phase itself is merely a convenient and simple way of imparting scalar phase (on light of a particular polarization state), one which moreover can be somewhat achromatic (since all wavelengths experience the angular orientations of the ridges identically). However, such structures are inherently polarization-sensitive -- while light of the desired circular polarization state picks up a phase that goes as $2\theta(x^{\prime},y^{\prime})$, the opposite handedness picks up a reciprocal phase of $-2\theta(x^{\prime},y^{\prime})$. The phase control is thus conjugate and not independently controllable for both circular polarization states. However, this can still be used advantageously when the desired function is conjugate between right- and left-handed circular polarization anyway, such as in the case of a circular polarization beamsplitter \cite{Bomzon2001} (which appears as opposite blazed gratings to oppositely circularly polarized light).

The space-variant Jones matrices implemented by this class of elements is given by

\begin{equation}
\label{eq:berry_phase_element}
    \boldsymbol{J}(x^{\prime},y^{\prime}) = \boldsymbol{R}\Big(\theta(x^{\prime},y^{\prime})\Big)
    \begin{pmatrix}
        i & 0 \\ 0 & -i
    \end{pmatrix}
    \boldsymbol{R}\Big(-\theta(x^{\prime},y^{\prime})\Big),
\end{equation}

with $\boldsymbol{R}(\xi)$ denoting a $2\times2$ rotation matrix for angle $\xi$. Eq. \ref{eq:berry_phase_element} essentially denotes a half-waveplate rotated in space with angle $\theta(x^{\prime},y^{\prime})$.

In retarder space, this is a 1D subspace of retarders with $\Delta = \pi$ with an eigenbasis that can take on any set of orthogonal linear polarization states (by rotation of the grating long axis, governed by $\theta$). Thus, within the Poincar\'e ball, this represents a belt along the equator. As the structure is rotated, overall phase represents one possible number, a point on the unit circle. This is depicted in Fig. \ref{fig:fig2}, second column from the right.

A notable application of these geometric phase elements has been in astronomy, where diamond structures have been developed for use in vector vortex geometric phase plates for direct imaging of exoplanets (see, e.g., \cite{Forsberg2013}.

\subsection{Present-day ``metasurfaces''}

Metasurfaces are the natural evolution of improved fabrication technology percolating into academic research environments in the last few decades. A typical metasurface architecture is an array of pillar-like structures with transverse profiles that are doubly mirror symmetric structures -- such as rectangles or ellipses. Each structure's length, width, and angular orientation can be arbitrarily varied over the aperture of the final device, effectively synthesizing a Jones matrix profile of the form:

\begin{multline}
    \label{eq:metasurface_element}
    \boldsymbol{J}(x^{\prime},y^{\prime}) = \\ \boldsymbol{R}\Big(\theta(x^{\prime},y^{\prime})\Big)
    \begin{pmatrix}
        e^{i\phi_x(x^{\prime},y^{\prime})} & 0 \\ 0 & e^{i\phi_y(x^{\prime},y^{\prime})}
    \end{pmatrix}
    \boldsymbol{R}\Big(-\theta(x^{\prime},y^{\prime})\Big).
\end{multline}

As is clear from Eq. \ref{eq:metasurface_element}, what are today called ``metasurfaces'' are a combination of all of the previous ideas discussed above into one. A single form-birefringent structure is modified in two-dimensions (instead of one, as in the grating groove-like form birefringence discussed above). The high refractive index contrast of these structures (made of materials such as silicon or TiO$_2$) with air permits each structure to be treated independently without inter-element crosstalk. This enables independent phase control of light linearly polarized along the long and short axes of the structures ($\phi_x$ and $\phi_y$), permitting control of overall phase $\phi$ and retardance $\Delta$ much like in the multi-level and form birefringent structures discussed above. However, the structures can also be rotated -- a degree of freedom inherited from the geometric phase-like elements described in the previous section -- enabling control over the polarization basis (angle $\chi$ in Eq. \ref{eq:normalized_jones}).

In retarder space, this corresponds to full coverage of $\phi$ (the entire unit circle), as well as an arbitrarily oriented linear polarization eigenbasis with arbitrary retardance (an area within the ``Poincar\'e ball'' that corresponds to the entire equitorial plane). This is a 3D subspace within retarder space, combining the degrees of freedom of the previously discussed technologies.

It should be explicitly noted that two conceptual advances led to the development of this enhanced polarization function by metasurface-like components. The first is the use of materials such as TiO$_2$ (in the visible) and amorphous Si (in the near infrared) which have high refractive index contrast with air. This permits tight confinement of light within these structures with minimized inter-element crosstalk, enabling metasurface structures to have features which can vary faster in space (compared to the wavelength) than the features in the old generations of diffractive elements discussed above which were (with some exceptions) fabricated from lower refractive index materials. One of the first predictions \cite{Lalanne1999} and demonstrations \cite{Lalanne1999d} of this capability was provided by Lalanne et al. in TiO$_2$, a demonstration that predated and anticipated the metasurface field over a decade later. The second conceptual advance is that in a metasurface, the rotation angle $\theta$ can vary arbitrarily from point-to-point, structure-to-structure, rather than in previous devices where $\theta$ varied continuously along ridges of material. This too is primarily owed to the high refractive index of contemporary metasurface structures which allows adjacent structures to be rotated at potentially totally different angles without crosstalk. It is unclear to us where this idea  first emerged in the diffractive optics literature --- starting in the early 2010s many works on metasurface-like components gravitated towards this natural degree-of-freedom of discrete rotating structures (notably, including Hasman et al. \cite{Shitrit2013, Lin2014} who had earlier pioneered the continuously rotating ridge geometric phase elements described in Sec. \ref{subsec:geometric_phase}).

The literature on metasurface polarization optics is too overwhelming in volume to be adequately summarized here, and thus we include just a few references (\cite{Rubin2021a} covers this in far more detail). The polarization-dependent behavior of metasurface-like elements has long been cited as among their chief advantages, but we contend that relatively few works in this field recognize their full possibility. Indeed, many (perhaps the majority) of works on metasurface polarization optics simply re-implement the sorts of devices that were earlier demonstrated with either form birefringent devices (in which independent phase profiles are imparted on orthogonal linear polarization states) or geometric phase (which impart a single phase profile on circularly polarized light). The true novelty of these metasurface optics is in the combination of both, as recognized by \cite{Arbabi2015, Mueller2017}. We revisit this point in the next section.

In describing the function of metasurface polarization optics, it may at first seem limiting that we have \emph{a priori} decided to describe simple structures such as rectangles or ellipses. After all, with the freedom afforded by lithographic fabrication the structures could take on most any shape in general (which has spawned a field of its own under the guise of ``inverse design''). However, it turns out that the level of polarization control described by Eq. \ref{eq:metasurface_element} is fundamental to a single layer of patterned nanostructures which possess mirror symmetry (which is usually a good approximation, given that the substrate -- frequently glass with $n\sim1.5$ -- only weakly breaks mirror symmetry). It can be shown from fundamental considerations that the Jones matrix of any single layer of nanostructures in transmission must be of the form of Eq. \ref{eq:metasurface_element} (see \cite{Menzel2010}, and a symmetry argument elaborated in \cite{Rubin2021a}, Sec. 4.7f) -- irrespective of the shape of the transverse profile of the layer. In other words, being a simple shape of neat symmetry (like a rectangle) is not required for the function described by Eq. \ref{eq:metasurface_element} to be complete. Eq. \ref{eq:metasurface_element} describes essentially the set of all unitary \emph{and} symmetric matrices \cite{Rubin2021, Rubin2021a}. In recent work, it has been shown that device consisting of \emph{two} aligned layers of metasurface pillar-like elements, in which two transformations of the form of Eq. \ref{eq:metasurface_element} are experienced in succession, is sufficient to realize \emph{any} unitary matrix, thus covering the entirety of retarder space \cite{Shi2022, Dorrah2025}.

It is then clear that metasurfaces represent the most general polarization control yet achieved with polarization-sensitive diffractive optics, combining past approaches which employed form birefringence with the freedom to spatially modulate the axis of that form birefringence, covering the largest yet-accessed region of retarder space. In the next section, we discuss new possibilities and perspectives this additional control enables, and some possible future directions.

\section{Implications and Conclusion}
\label{sec:implications_and_conclusion}

\subsection{The rise of Jones matrix Fourier optics}

Fig. \ref{fig:fig2} and the previous section provide a technical argument for why metasurface optics represent a more general form of space-variant polarization control than similar diffractive optics technologies of decades past. However, we further argue that this difference, which may at first seem to be merely categorical in nature, requires metasurfaces to be treated in a altogether more sophisticated way. This has enabled new technologies and a more powerful view of the role of light's polarization in diffractive optics more generally.

What is notable about the ``older'' generation of polarization-sensitive diffractive optics is that their simplicity allows these to be understood and designed without a fully polarization-dependent mathematical treatment. For one, the simple form birefringent or multi-level birefringent devices (as described by Eq. \ref{eq:pol_control_fbh}) impart independent phase profiles on $x$ and $y$ polarized light. The problem of analyzing and designing these is then immediately reduced to two completely independent problems of scalar diffraction. The case of geometric phase devices is even simpler -- provided the designer always operates with circularly polarized light, these devices impart a scalar phase profile and can be designed and understood (as conventionally used) in a way that is no different from a ``normal'' diffractive optical element described without reference to polarization at all.

On the other hand, the function of metasurface-like devices as given by Eq. \ref{eq:metasurface_element} \emph{cannot} in general be reduced to one or two scalar diffraction problems. This means that metasurface polarization optics are fundamentally polarization-dependent devices in a way that is unlike their predecessors. The design and analysis of these devices must remain at the level of the space-variant Jones matrix, which necessitates an explicit consideration of the polarization state of light involved. This is complex and not readily intuitable which is why, as remarked above, much work in metasurface polarization optics has simply re-implemented simple $x$/$y$ form birefringent or geometric phase devices that had long before been demonstrated.

Full treatment of this complexity requires a merger of the Jones calculus with the linear systems theory of Fourier optics \cite{Rubin2019, Rubin2021a}. In this way, the diffraction of light by a metasurface can be specified for all possible illuminating polarization states \emph{simultaneously}. In Jones matrix Fourier optics, the space-variant Jones matrix implemented by a metasurface can itself be Fourier transformed element-wise, yielding a second Jones matrix transformation which encodes the polarization-dependence of the far-field.

The earlier technologies above can also be seen through this lens, albeit in a way that is trivial and does not require a full matrix formalism. Metasurfaces, however, require this in general -- and are the first\footnote{There are some caveats to this, as qualified in \cite{Rubin2021a}.} technology sufficiently complex to require this treatment. This has opened a new direction of classical optics technology and research, and has enabled a significantly expanded design space for polarization optics. For instance, Jones matrix Fourier optics and metasurfaces permit the realization of massively multitasking polarization optics, in which many polarization functions can be combined into one optical element, accessed independently with, for instance, angular position in the far field \cite{Rubin2021} or even along the propagation direction \cite{Dorrah2021}. Among the more interesting of these elements are metasurface devices in which light is sorted in a parallel manner between different angular channels according to its polarization state, either over an entire holographic image \cite{Rubin2021} or discrete diffraction orders \cite{Rubin2019}. These devices enable the simplification of polarimetric imaging technology \cite{Rubin2022}.

\subsection{Future outlook}

Polarization-dependent diffractive optics, may see broad application in scientific instrumentation and perhaps consumer devices, especially given the added functionality, fabrication maturity, and flexibility of polarization control afforded by metasurfaces in particular. Research and applications in these directions are only just beginning. Rather than focus on new technologies here, however, in the spirit of this Perspective we conclude instead with some very brief remarks on future research in this area that may enable new optical functions.

In our opinion, the consequences of Jones matrix Fourier optics for optical systems are not yet fully understood. The Fourier optics of linear optical systems can be fully revisited from a matrix perspective with the knowledge that devices exist (metasurfaces) to implement spatial, polarization-dependent transfer functions. We note that this is a pursuit which has already attracted some interest \cite{Wang2024}. This may additionally hold consequences for quantum optics, where polarization is commonly used as an entangled degree-of-freedom. Metasurfaces have been applied here \cite{Wang2018}, albeit in ways that are, so far, quite simplistic.

Recent years have additionally seen significant interest in multi-layer metasurface devices \cite{Shi2022, Mirzapourbeinekalaye2022, Dorrah2025}. From a functional perspective, bilayer metasurface devices can enable full unitary matrix control of the local transfer function \cite{Shi2022}, relieving some constraints on the far-field transfer functions posed by single layer devices. Moreover, we believe that the freedom provided by these devices may aide in achromatizing the function of metasurface polarization optics.

%\tableofcontents

\bibliographystyle{ieeetr}

\begin{thebibliography}{10}

\bibitem{Lalanne2017}
P.~Lalanne and P.~Chavel, ``Metalenses at visible wavelengths: past, present, perspectives,'' {\em Laser \& Photonics Reviews}, vol.~11, no.~3, p.~1600295.

\bibitem{Banerji2019}
S.~Banerji, M.~Meem, A.~Majumder, F.~G. Vasquez, B.~Sensale-Rodriguez, and R.~Menon, ``Imaging with flat optics: metalenses or diffractive lenses?,'' {\em Optica}, vol.~6, no.~6, pp.~805--810.

\bibitem{Lalanne2023}
P.~Lalanne and P.~Chavel, ``On the prehistory of optical metasurfaces,'' {\em Photoniques}, vol.~119, pp.~41--45, 2023.

\bibitem{Rubin2021a}
N.~A. Rubin, Z.~Shi, and F.~Capasso, ``Polarization in diffractive optics and metasurfaces,'' {\em Adv. Opt. Photon.}, vol.~13, pp.~836--970, 2021.

\bibitem{Chipman2019}
R.~A. Chipman, W.-S.~T. Lam, and G.~Young, {\em Polarized Light and Optical Systems}.
\newblock CRC Press, 2019.

\bibitem{Ford1993}
J.~E. Ford, F.~Xu, K.~Urquhart, and Y.~Fainman, ``Polarization-selective computer-generated holograms,'' {\em Optics Letters}, vol.~18, pp.~456--458, 1993.

\bibitem{Xu1995}
F.~Xu, J.~E. Ford, and Y.~Fainman, ``Polarization-selective computer-generated holograms: design, fabrication, and applications,'' {\em Appl. Opt.}, vol.~34, pp.~256--266, 1995.

\bibitem{Xu1996a}
F.~Xu, R.~chung Tyan, Y.~Fainman, and J.~E. Ford, ``Single-substrate birefringent computer-generated holograms,'' {\em Optics Letters}, vol.~21, pp.~516--518, 1996.

\bibitem{Krishnamoorthy1997}
A.~V. Krishnamoorthy, F.~Xu, J.~E. Ford, and Y.~Fainman, ``Polarization-controlled multistage switch based on polarization-selective computer-generated holograms,'' {\em Applied Optics}, vol.~36, pp.~997--1010, 1997.

\bibitem{Richter1995}
I.~Richter, P.-C. Sun, F.~Xu, and Y.~Fainman, ``Design considerations of form birefringent microstructures,'' {\em Applied Optics}, vol.~34, p.~2421, 1995.

\bibitem{Xu1996}
F.~Xu, R.~chung Tyan, P.~chen Sun, Y.~Fainman, C.~cheng Cheng, and A.~Scherer, ``Form-birefringent computer-generated holograms,'' {\em Appl. Opt.}, vol.~21, pp.~1513--1515, 1996.

\bibitem{Tyan1997}
R.-C. Tyan, A.~A. Salvekar, H.-P. Chou, C.-C. Cheng, A.~Scherer, P.-C. Sun, F.~Xu, and Y.~Fainman, ``Design, fabrication, and characterization of form-birefringent multilayer polarizing beam splitter,'' {\em Journal of the Optical Society of America A}, vol.~14, pp.~1627--1636, 1997.

\bibitem{Lopez1998}
A.~G. Lopez and H.~G. Craighead, ``Wave-plate polarizing beam splitter based on a form-birefringent multilayer grating,'' {\em Optics Letters}, vol.~23, pp.~1627--9, 1998.

\bibitem{Lopez2001}
A.~G. Lopez and H.~G. Craighead, ``Subwavelength surface-relief gratings fabricated by microcontact printing of self-assembled monolayers,'' {\em Appl. Opt.}, vol.~40, no.~13, pp.~2068--2075.

\bibitem{Aoyama1991}
S.~Aoyama and T.~Yamashita, ``Grating beam splitting polarizer using multilayer resist method,'' in {\em International Conference on the Application and Theory of Periodic Structures} (J.~M. Lerner and W.~R. McKinney, eds.), vol.~1545, pp.~241 -- 250, International Society for Optics and Photonics, SPIE, 1991.

\bibitem{Born1999}
M.~Born and E.~Wolf, {\em Principles of Optics}.
\newblock Pergamon, 1999.

\bibitem{Rytov1956}
S.~M. Rytov, ``Electromagnetic properties of a finely stratified medium,'' {\em Soviet Physics JETP}, vol.~2, pp.~466--475, 1956.

\bibitem{Bomzon2002c}
Z.~Bomzon, G.~Biener, V.~Kleiner, and E.~Hasman, ``Space-variant pancharatnam–berry phase optical elements with computer-generated subwavelength gratings,'' {\em Optics Letters}, vol.~27, pp.~1141--1143, 2002.

\bibitem{Hasman2005}
E.~Hasman, G.~Biener, A.~Niv, and V.~Kleiner, ``Space-variant polarization manipulation,'' {\em Progress in Optics}, vol.~47, pp.~215--289, 2005.

\bibitem{Levy2004}
U.~Levy, C.-H. Tsai, L.~Pang, and Y.~Fainman, ``Engineering space-variant inhomogeneous media for polarization control,'' {\em Optics Letters}, vol.~29, p.~1718, 2004.

\bibitem{Tsai2006}
C.-H. Tsai, U.~Levy, L.~Pang, and Y.~Fainman, ``Form-birefringent space-variant inhomogeneous medium element for shaping point-spread functions,'' vol.~45, no.~8, pp.~1777--1784.

\bibitem{Bomzon2001}
Z.~Bomzon, V.~Kleiner, and E.~Hasman, ``Pancharatnam-berry phase in space-variant polarization-state manipulations with subwavelength gratings.,'' {\em Optics Letters}, vol.~26, pp.~1424--1426, 2001.

\bibitem{Forsberg2013}
P.~Forsberg and M.~Karlsson, ``High aspect ratio optical gratings in diamond,'' {\em Diamond and Related Materials}, vol.~34, pp.~19--24, 2013.

\bibitem{Lalanne1999}
P.~Lalanne, ``Waveguiding in blazed-binary diffractive elements,'' {\em Journal of the Optical Society of America A}, vol.~16, no.~10, p.~2517, 1999.

\bibitem{Lalanne1999d}
P.~Lalanne, S.~Astilean, P.~Chavel, E.~Cambril, and H.~Launois, ``{Design and fabrication of blazed binary diffractive elements with sampling periods smaller than the structural cutoff},'' {\em Journal of the Optical Society of America A}, vol.~16, no.~5, p. 1143, 1999.

\bibitem{Shitrit2013}
N.~Shitrit, I.~Yulevich, E.~Maguid, D.~Ozeri, D.~Veksler, V.~Kleiner, and E.~Hasman, ``Spin-optical metamaterial route to spin-controlled photonics,'' {\em Science}, vol.~340, no.~6133, pp.~724--726, 2013.

\bibitem{Lin2014}
D.~Lin, P.~Fan, E.~Hasman, and M.~L. Brongersma, ``Dielectric gradient metasurface optical elements,'' {\em Science}, vol.~345, pp.~298--302, 2014.

\bibitem{Arbabi2015}
A.~Arbabi, Y.~Horie, M.~Bagheri, and A.~Faraon, ``Dielectric metasurfaces for complete control of phase and polarization with subwavelength spatial resolution and high transmission,'' {\em Nature Nanotechnology}, vol. 10, no. 11, pp. 937-43, 2015.

\bibitem{Mueller2017}
J.~P.~B. Mueller, N.~A. Rubin, R.~C. Devlin, B.~Groever, and F.~Capasso, ``Metasurface polarization optics: independent phase control of arbitrary orthogonal states of polarization,'' {\em Physical Review Letters}, vol.~118, p.~113901, 2017.

\bibitem{Menzel2010}
C.~Menzel, C.~Rockstuhl, and F.~Lederer, ``Advanced jones calculus for the classification of periodic metamaterials,'' {\em Physical Review A}, vol.~82, p.~053811, 2010.

\bibitem{Rubin2021}
N.~A. Rubin, A.~Zaidi, A.~H. Dorrah, Z.~Shi, and F.~Capasso, ``Jones matrix holography with metasurfaces,'' {\em Science Advances}, vol.~7, p.~eabg7488, 2021.

\bibitem{Shi2022}
Z.~Shi, N.~A. Rubin, J.-S. Park, and F.~Capasso, ``Nonseparable polarization wavefront transformation,'' {\em Physical Review Letters}, vol.~129, p.~167403.

\bibitem{Dorrah2025}
A.~H. Dorrah, J.-S. Park, A.~Palmieri, and F.~Capasso, ``Free-standing bilayer metasurfaces in the visible,'' {\em Nature Communications}, vol.~16, no.~1, p.~3126, 2025.

\bibitem{Rubin2019}
N.~A. Rubin, G.~D'Aversa, P.~Chevalier, Z.~Shi, W.~T. Chen, and F.~Capasso, ``Matrix Fourier optics enables a compact full-Stokes polarization camera,'' {\em Science}, vol.~365, p.~eaax1839, 2019.

\bibitem{Dorrah2021}
A.~H. Dorrah, N.~A. Rubin, A.~Zaidi, M.~Tamagnone, and F.~Capasso, ``Metasurface optics for on-demand polarization transformations along the optical path,'' {\em Nature Photonics}, vol.~15, pp.~287--296, 2021.

\bibitem{Rubin2022}
N.~A. Rubin, P.~Chevalier, M.~Juhl, M.~Tamagnone, R.~Chipman, and F.~Capasso, ``Imaging polarimetry through metasurface polarization gratings,'' {\em Optics Express}, vol.~30, p.~9389, 3 2022.

\bibitem{Wang2024}
W.~Wang, ``Matrix-based fourier analysis of matrix signals and systems for polarization optics,'' {\em Journal of the Optical Society of America A}, vol.~41, no.~10, pp.~1969--1978.

\bibitem{Wang2018}
K.~Wang, J.~G. Titchener, S.~S. Kruk, L.~Xu, H.~Pin Chung, M.~Parry, I.~I. Kravchenko, Y.~Hung Chen, A.~S. Solntsev, Y.~S. Kivshar, D.~N. Neshev, and A.~A. Sukhorukov, ``Quantum metasurface for multiphoton interference and state reconstruction,'' {\em Science}, vol.~361, pp.~1104--1108, 2018.

\bibitem{Mirzapourbeinekalaye2022}
B.~Mirzapourbeinekalaye, A.~McClung, and A.~Arbabi, ``General lossless polarization and phase transformation using bilayer metasurfaces,'' {\em Adv. Opt. Mat.}, vol.~10, no.~11, p.~2102591.

\bibitem{Gaylord1981}
T.~K. Gaylord and M.~Moharam, ``Thin and thick gratings: terminology clarification,'' {\em Appl. Opt.}, vol.~20, no.~19, pp.~3271--3273, 1981.

\bibitem{Gaylord1985}
T.~K. Gaylord and M.~Moharam, ``Analysis and applications of optical diffraction by gratings,'' {\em Proc. IEEE}, vol.~73, no.~5, pp.~894--937, 1985.

\end{thebibliography}

\end{document}